# Polynomial Weights or Generalized Geometric Weights: Yet Another Scheme for Assigning Credits to Multiple Authors

Ash Mohammad Abbas
Department of Computer Engineering
Aligarh Muslim University
Aligarh - 202002, India
am.abbas.ce@amu.ac.in

June 4, 2018

**Abstract**

Devising a weight assignment policy for assigning credits to multiple authors of a manuscript is a challenging task. In this paper, we present a scheme for assigning credits to multiple authors that we call a *polynomial weight assignment scheme*. We compare our scheme with other schemes proposed in the literature.

## 1 Introduction

Assigning weights to multiple authors of a paper is a challenging task. The challenge comes from the fact that the conventions followed among different areas of research are different, and there is no universally agreed upon policy for sharing credits of a multiauthored paper. The reason is that the conventions followed among research groups might not depend only on the academic factors, sometimes, these convention might also depend on the social, economic, regional, and scientific factors. Actually, a weight assignment scheme can only generate weights. How judiciously the scheme is applied is the responsibility of the one who is evaluating the quality of research produced by an author or a set of authors. How the weights generated by a weight assignment policy can be made in the conformance of the conventions followed by a research group is partly described in [3]. We here focus on the schemes that generate weights (and not that are applied to a particular research field).

Many researchers focused on weight assignment schemes for sharing credits among multiple authors of a paper. Addressing multiple authorship mathematically dates back to [4]. In [2], a weight assignment scheme called *generalized*

*linear weights* is presented where the weights are a generalized version of the *proportional* or *arithmetic* weights [5], [8]. Other schemes proposed in the literature include *fractional* or *equal* [6], [7], *geometric* [9], *harmonic* [11], [19]. We have described the related work in the later part of this paper.

In this paper, we present a weight assignment policy that we call polynomial weights for sharing credits among multiple authors of a paper. We then compare the polynomial weights with other types of weights proposed in the literature such as *equal weights* and *geometric weights*.

The rest of the paper is organized as follows. In section 2, we present polynomial weights. In section 3, we compare the proposed weights with other types of weights presented in the literature. Section 4 contains results and discussion. Section 5 is for related work. The last section contains conclusion and future works.

## 2  Polynomial Weights

We define a weight assignment scheme that we call *Polynomial Weights: Type-I*.

**Definition 1** (Plynomial Weights: Type-I). *Let $x \leq 1$ be a variable that we call* weight control parameter. *Let there be $k$ authors of a paper. The weight of the $j$th author of the paper is as follows.*

$$w_j = \begin{cases} 1 & \text{for } k = 1 \\ \frac{x^{j-1}}{\sum_{i=1}^{k} x^{i-1}} & \text{for } k > 1, 1 \leq j \leq k \end{cases}.$$

**Example 1.** *Let there be 3 authors of a paper. The weights of the authors are as follows.*

$$\begin{aligned} w_1 &= \frac{1}{1+x+x^2} \\ w_2 &= \frac{x}{1+x+x^2} \\ w_3 &= \frac{x^2}{1+x+x^2}. \end{aligned} \quad (1)$$

Similarly, we can write weights for any number of authors of a paper. We now state the following lemma.

**Lemma 1.** *The equation (1) is a weight assignment policy.*

*Proof.* To prove that (1) is a weight assignment policy, we need to show that (i) $\sum_{i=j}^{k} w_j = 1$, and that (ii) the weights, $w_j$, are in decreasing order with increasing $j$.

Note for $k = 1$, the summation of weights is trivially equal to 1. For $k \geq 2$,

we have,

$$\begin{aligned}
\sum_{j=1}^{k} w_j &= \frac{1}{\sum_{i=1}^{k} x^{i-1}} \left\{ x^0 + x^1 + ... + x^{k-1} \right\} \\
&= \frac{1}{\sum_{i=1}^{k} x^{i-1}} \left\{ \sum_{j=1}^{k} x^{j-1} \right\} \\
&= 1.
\end{aligned} \quad (2)$$

To show the second condition, we find the derivative of $w_j$ with respect to $x$. We get,

$$\frac{dw_j}{dj} = \frac{x^{j-1} \ln x}{\sum_{i=1}^{k} x^{i-1}}. \quad (3)$$

For $x < 1$, $\ln x = -ve$. It implies that

$$\frac{dw_j}{dj} = -ve. \quad (4)$$

It infers that the weights decrease from the first to the last author. Therefore, (1) represents a weight assignment policy. □

We now state another lemma that tells how polynomial weights are related to the equal weights.

**Lemma 2.** *The polynomial weights for $x = 1$ are same as equal weights.*

Putting $x = 1$ in the Example 1, weights for $k = 3$ authors are $w_1 = \frac{1}{3}$, $w_2 = \frac{1}{3}$, and $w_3 = \frac{1}{3}$.

We now define polynomial weights when the parameter $x > 1$. We call these weights as *Polynomial Weights of Type II*.

**Definition 2** (Plynomial Weights: Type-II)**.** *Let $x \geq 1$ be a variable that we call* weight control parameter. *Let there be $k$ authors of a paper. The weight of the $j$th author of the paper is as follows.*

$$w_j = \begin{cases} 1 & k = 1 \\ \frac{x^{k-j}}{\sum_{i=1}^{k} x^{i-1}} & k > 1, 1 \leq j \leq k \end{cases}.$$

We now consider an example to better understand *Polynomial Weights of Type-2*.

**Example 2.** *Let the number of authors, $k = 3$, then the weights of author 1 through author 3 are as follows.*

$$\begin{aligned}
w_1 &= \frac{x^2}{1 + x + x^2} \\
w_2 &= \frac{x}{1 + x + x^2} \\
w_3 &= \frac{1}{1 + x + x^2}.
\end{aligned} \quad (5)$$

The difference is that the expressions for Type-II weights are in the reverse order of Type-I weights. We now have the following lemma.

**Lemma 3.** *The equation (5) is a weight assignment scheme.*

*Proof.* Again, to prove that (5) is a weight assignment scheme, we need to show that (i) $\sum_{i=j}^{k} w_j = 1$, and that (ii) the weights, $w_j$, are in decreasing order with increasing $j$.

Note for $k = 1$, the summation of weights is trivially equal to 1. Also, for $k \geq 2$, the summation of weights is equal to 1. For the second condition, we have,

$$\frac{dw_j}{dj} = \frac{x^{k-j} \ln x.(-1)}{\sum_{i=1}^{k} x^{i-1}}$$
$$\frac{dw_j}{dj} = -\frac{x^{k-j} \ln x}{\sum_{i=1}^{k} x^{i-1}}. \quad (6)$$

For $x \geq 1$, $\ln x$ is +ve. Therefore, $\frac{dw_j}{dj} = -ve$. It implies that the weights decrease from the first to the last author. As a result, (5) is a weight assignment scheme. □

In what follows, we compare polynomial weights with other types of weights.

## 3 Comparison with Other Weight Assignment Schemes

In this section, we compare polynomial weights with other types of weights such as *equal weights* and *geometric weights*.

### 3.1 Comparison Between Polynomial Weights and Equal Weights

We state the following lemma that tells how polynomial weights are related to the equal weights.

**Lemma 4.** *The polynomial weights for $x = 1$ are same as equal weights.*

*Proof.* For $x = 1$, we have either from (1) or from (5),

$$w_j = \frac{1}{\sum_{i=1}^{k} 1}$$
$$= \frac{1}{k} \quad (7)$$

which gives the weight of each author under equal weight assignment scheme. □

We can also understand it by putting $x = 1$ either in the Example 1 or in Example 2, weights for $k = 3$ authors are $w_1 = \frac{1}{3}$, $w_2 = \frac{1}{3}$, and $w_3 = \frac{1}{3}$.

We now compare the polynomial weights with geometrical weights.

## 3.2 Comparison Between Polynomial Weights and Geometrical Weights

Let there be $k$ authors of a paper, the weight of $j$th author under *geometric weight assignment scheme* is given by the following expression.

$$w_j = \frac{2^{k-j}}{2^k - 1}. \tag{8}$$

On the other hand, the denominator in the R.H.S. of (5), which is the expression for *Polynomial Weights: Type-II*, can be written as follows.

$$\sum_{i=1}^{k} x^{i-1} = \frac{x^k - 1}{x - 1}. \tag{9}$$

Using (5) and (9), we have,

$$w_j = \frac{x^{k-j}(x-1)}{x^k - 1}. \tag{10}$$

Now, putting $x = 2$ in (10), we have,

$$w_j = \frac{2^{k-j}}{2^k - 1}$$

which is nothing but (8), an expression for geometric weights. Actually, (5) is a generalized expression for geometric weights where $x > 1$. Similarly, one can say that (1) is a generalized expression for geometrical weights with $x < 1$. As mentioned earlier, for $x = 1$, the polynomial weights are nothing but the equal weights. Therefore, we can say that the polynomial weights as given by (1) and (5) are *generalized geometrical weights*. Specifically, *Polynomial Weights: Type-I* as given by (1) are generalized geometrical weights for $x < 1$ and the *Polynomial Weights: Type-II* as given by (5) are generalized geometrical weights for $x > 1$.

## 4 Analysis and Discussion

In this section, we analyze polynomial weights. We begin our analysis with the following Theorem.

**Theorem 1.** *Let $k$ be the number of authors of a paper. The polynomial wights with $x < 1$ (i.e. Polynomial Weights: Type-I) comes out to be the weights with $x > 1$ (i.e. Polynomial Weights: Type-II) and vice versa, when $x$ is replaced with $\frac{1}{x}$.*

*Proof.* Let us start with *Polynomial Weights: Type-I*. The weight of $j$th author under polynomial weight assignment policy, for $x < 1$ is,

$$w_j = \frac{x^{j-1}(1-x)}{1-x^k}.$$

Let $x = \frac{1}{q}$. The weight of $j$th author under *Polynomial Weights: Type-I*, is given by,

$$\begin{aligned} w_j &= \frac{\left(\frac{1}{q}\right)^{j-1}\left(1-\frac{1}{q}\right)}{1-\frac{1}{q}^k} \\ &= \frac{\frac{1}{q^{j-1}}\frac{q-1}{q}}{1-\left(\frac{1}{q}\right)^k} \\ &= \frac{q-1}{q^j}\frac{q^k}{q^k-1} \\ &= \frac{q^{k-j}(q-1)}{q^k-1}. \end{aligned} \quad (11)$$

Note that this is an expression for the weight of $j$th author under *Polynomial Weights: Type-II* as defined by (10) for $q = \frac{1}{x}$. Similarly, we can show that Type-II weights can become Type-I weights. □

We now state a theorem that relates the weights of the first and last authors under polynomial weight assignment policy.

**Theorem 2.** *Let $k$ be the number of authors of a paper, and $k >> 1$. The ratio of weights of the first and the last authors of the paper under polynomial weight assignment policy is as follows.*

$$\frac{w_1}{w_k} \approx \begin{cases} x^{k-1} & x > 1 \\ \left(\frac{1}{x}\right)^{k-1} & x < 1. \end{cases}$$

*Proof.* Let us first consider $x > 1$. For $x > 1$ and $k >> 1$, $x^k >> 1$, therefore, $x^k - 1 \approx x^k$. Using (10), the expression for the weight of author $j$ can be written as follows.

$$\begin{aligned} w_j &\approx \frac{x^{k-j}(x-1)}{x^k} \\ &\approx x^{-j}(x-1). \end{aligned} \quad (12)$$

Using (12), the weights of the first author and the last author are given by,

$$\begin{aligned} w_1 &\approx \frac{x-1}{x} \\ w_k &\approx \frac{x-1}{x^k}. \end{aligned} \quad (13)$$

The ratio of the weights of the first author and the last author is,

$$\frac{w_1}{w_k} \approx \frac{x^k}{x}$$
$$\approx x^{k-1}. \tag{14}$$

Let us now consider the weight of $j$th author when $x < 1$. For $x < 1$ and $k >> 1$, we have, $1 - x^k \approx 1$. Therefore,

$$w_j = x^{j-1}(1-x). \tag{15}$$

Using (15), the weights of the first author and the last author are given by,

$$w_1 \approx 1-x$$
$$w_k \approx x^{k-1}(1-x). \tag{16}$$

The ratio of the weights of first author and the last author is,

$$\frac{w_1}{w_2} \approx \frac{1}{x^{k-1}}$$
$$\approx \left(\frac{1}{x}\right)^{k-1}. \tag{17}$$

□

Consider for example $k = 3$. For $x = 2$, the weight of the first author is 4 times the weight of the last author (as given by 14). For $x = 0.5$, the weight of the first author is again 4 times the weight of the last author (as given by (17)).

Table 1 shows the weights of individual authors for different values of $k$. Note that the weights of authors for a given value of $k$, is the same for $x = 2$ and for $x = 0.5$. This confirms the statement of Theorem 1. We would like to mention that these weights are the same as the geometric weights as given by (8).

Figure 1 shows the weights of the first and the last authors as a function of the number of authors of a paper. We observe that under polynomial weight assignment (or generalized geometric weights) the weights of the first author and the last authors decrease with an increase in the number of authors of a paper. However, the weight of the last author decreases more rapidly as compared to the weight of the first author. Specifically, the weight of the first author decreases linearly, however, the weight of the last author decreases exponentially.

## 5 Related Work

Many researchers have focused on the problem of sharing credits among multiple authors of a paper from different perspectives. Mathematically, addressing multiple authorship dates back to [4]. In [15], the failure of equal weight assignment scheme to multiple authors of a paper is described. A study regarding the money value of citations in single authored and multi-authored articles appeared in [16]. A method for sharing credits among multiple authors called *Correct Credit Distribution* (CCD) is proposed in [17], where the weights are called *Corrected Contribution Scores*, and are based on the minimum and the maximum contribution scores of authors. In [18], it has been discussed that in multi-authored papers, the roles of authors should be defined prior to allocating credits of authorship. A review of the research on alphabetized ordering of authors in the field of economics, physics, and information science is presented in [20], in which a study is carried out for a period of past 30 years ( specifically, from $1978 - 2007$). The outcome of the study shows that there is a significant drop in alphabetized ordering in the field of information science, and a significant increase in the field of economics. The state of the art in publication counting is described in [10].

Table 1: Number of authors and polynomial weights of individual authors (for $x = 2$ or for $x = 0.5$.

| Number of Authors | $w_1$ | $w_2$ | $w_3$ | $w_4$ | $w_5$ | $w_6$ | $w_7$ | $w_8$ | $w_9$ | $w_{10}$ |
|---|---|---|---|---|---|---|---|---|---|---|
| 1 | 1 | | | | | | | | | |
| 2 | $\frac{2}{3}$ | $\frac{1}{3}$ | | | | | | | | |
| 3 | $\frac{4}{7}$ | $\frac{2}{7}$ | $\frac{1}{7}$ | | | | | | | |
| 4 | $\frac{8}{15}$ | $\frac{4}{15}$ | $\frac{2}{15}$ | $\frac{1}{15}$ | | | | | | |
| 5 | $\frac{16}{31}$ | $\frac{8}{31}$ | $\frac{4}{31}$ | $\frac{2}{31}$ | $\frac{1}{31}$ | | | | | |
| 6 | $\frac{32}{63}$ | $\frac{16}{63}$ | $\frac{8}{63}$ | $\frac{4}{63}$ | $\frac{2}{63}$ | $\frac{1}{63}$ | | | | |
| 7 | $\frac{64}{127}$ | $\frac{32}{127}$ | $\frac{16}{127}$ | $\frac{8}{127}$ | $\frac{4}{127}$ | $\frac{2}{127}$ | $\frac{1}{127}$ | | | |
| 8 | $\frac{128}{255}$ | $\frac{64}{255}$ | $\frac{32}{255}$ | $\frac{16}{255}$ | $\frac{8}{255}$ | $\frac{4}{255}$ | $\frac{2}{255}$ | $\frac{1}{255}$ | | |
| 9 | $\frac{256}{511}$ | $\frac{128}{511}$ | $\frac{64}{511}$ | $\frac{32}{511}$ | $\frac{16}{511}$ | $\frac{8}{511}$ | $\frac{4}{511}$ | $\frac{2}{511}$ | $\frac{1}{511}$ | |
| 10 | $\frac{512}{1023}$ | $\frac{256}{1023}$ | $\frac{128}{1023}$ | $\frac{64}{1023}$ | $\frac{32}{1023}$ | $\frac{16}{1023}$ | $\frac{8}{1023}$ | $\frac{4}{1023}$ | $\frac{2}{1023}$ | $\frac{1}{1023}$ |

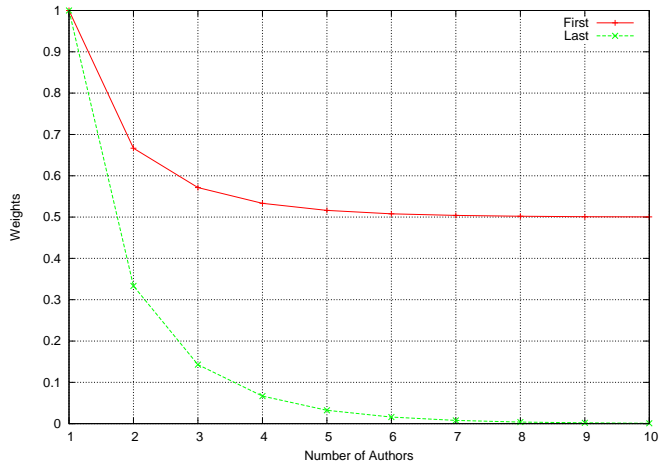

Figure 1: The weights of the first and last authors as a function of the number of authors for $x = 2$ or $x = \frac{1}{2}$.

Schemes for weight assignment to multiple authors are proposed by a number of researchers. These schemes include *proportional* (or *arithmetic*) [5] [8], *geometric* [9], and *fractional* (or *equal*) [6], [7]. A study of the linear growth in the percentage of equal first authors and corresponding authors is presented in [19]. The works presented in [19] and [11] favor the harmonic weight assignment to individual authors of the paper. In [2], we presented a generalized linear weights for assigning credits to multiple authors. How the weights generated by a weight assignment policy can be plugged in to achieve an index is described in [1].

Note that the conventions of how to share the credits might be different among different research disciplines. A weight assignment policy can only generate the weights according predefined assumptions. How to apply the weights generated by a weight assignment policy is the responsibility of the one who wishes to evaluate the quality of research produced by an author or a set of authors according to the domain of research. How the weights generated by a weight assignment policy can be made to conform to the conventions followed in a research discipline has been addressed in [3].

In this paper, we have described polynomial weights for sharing credits among multiple authors of a paper. These polynomial weights are nonlinear weights i.e. the weights authors, for a given number of authors, do not decrease linearly. The weights can be varied depending upon the weight control parameter, $x$. For $x = 1$, the polynomial weights are the same as the *equal weights*. For $x = 2$, the polynomial weights are the same as what is called *geometric weights* in the literature. For this reason, we call the polynomial weights as *generalized geometric weights*.

## 6 Conclusion

In this paper, we presented a weight assignment scheme that we called *polynomial weights* for assigning credits to multiple authors of a paper in decreasing order from the first author to the last author. However, if one wishes, a slight modification of the weights can provide weights in increasing order. The polynomial weights can be varied depending upon the *weight control parameter*. We have shown that for the weight control parameter, $x = 1$, the polynomial weights become *equal weights* for all authors of the paper. In other words, for $x = 1$, the credits among each author are shared equally. Further, for $x \neq 1$, the the polynomial weights resemble the geometric weights. We, therefore, call polynomial weights for $x \neq 1$ as *generalized geometric weights*. Further, validation of the proposed weight assignment scheme for different ressearch fields forms the future work.